\documentstyle[12pt, aaspp]{article} 
\begin{document}
\title{The C/CO Ratio Problem:\\ Chemical Effects of Turbulence}

\author{Taoling Xie}
\affil{Laboratory for Millimeter-wave Astronomy\\
           Department of Astronomy, University of Maryland\\
	   College Park, MD 20742; email: tao@astro.umd.edu}
{\bf ``CO: Twenty Five Years of Millimeter Wave Spectroscopy", Proceedings of the 170th Symposium of the International Astronomical Union, Tucson, Arizona, 1995 May 29-June 5, Eds. W.B.Latter et al, (Kluwer: Dordrecht) }
\begin{abstract}
We further discuss the suggestion that the chemistry in a photon-dominated 
region is coupled to that in the UV-shielded region behind it by 
turbulent transport processes. 
In addition to transport time-scales, we discuss 
why MHD waves/turbulence will likely cause transport instead of prohibiting it.
\end{abstract}

\section{On the C/CO Ratio Problem}

The study of interstellar chemistry has so far focused mainly on two 
physical regimes; regions subject to abundant ultra-violet photons, and 
regions which are shielded from UV photons. The 
former is often referred to as photon-dominated regions (so-called PDRs) where 
the chemistry is dominated by photon-related reactions, and the latter is 
often thought to be cosmic ray-dominated regions (hereafter CDRs) where cosmic 
rays sustain the ion-neutral reaction scheme.

The observed large $C/CO$ ratio ($\sim 0.1$) from molecular clouds has been a 
major challenge to theoreticians working in either regimes (cf. Sorrell 1992).
One common opinion in the field is that the clumpiness of a cloud
allows an enhanced UV penetration into the cloud interior which will dissociate
the molecules and create many small photon-dominated regions (PDRs) on the 
clump surfaces. One severe difficulty in this picture, however, 
seems to be the large observed abundances of many large organic molecules. 
Since large organic 
molecules do not survive UV photons in PDRs, the observed large organic 
molecules must be deep inside the clumps, which requires the co-existence 
of abundant neutral carbon (Chi\`{e}ze et al 1991). The co-volume existence of 
neutral carbon with CO is strongly indicated by the similarity of their line 
profiles (cf. Walker et al 1993) and the tight correlation between 
their emission intensities (cf. Tauber et al 1995).

\section{On Turbulent Transport and Its Time-scales}

What feasible mechanism may bring along a lot of neutral carbon deep inside
each clump without those UV photons that also destroy large organic molecules ?
An easy choice seems to be particle transport, which is known to occur 
universally in random gas motions. Xie, Allen \& Langer (1995, 
hereafter XAL) has formulated
the idea in the framework of the Prandtl mixing-length theory for turbulent 
diffusion and has found that the transport may give rise to a C/CO ratio in accord
with observations. 

Transport of particles occurs as a result of 
molecular diffusion. If there is a wind, convection or
a large-scale random component in gas motions, then the transport due to any 
of these processes is likely to dominate over molecular diffusion due to the much 
larger spatial scales and amplitudes over which these processes occur in 
interstellar clouds.  In general, 
random motions cause fluctuations of a physical quantity in a non-uniform 
multi-component gas. 
Defining the fractional abundances $f_{i}=n_{i}/n(H_{2})$ for species $i$, the transport flux can be written as 
(cf.\ Colegrove et al 1966; XAL), 
$\phi_{i} [ cm^{-2}\; s^{-1}] = n(H_{2})<V_{t}\delta f_{i}> = -Kn(H_{2})\frac{df_
{i}}{dz} = -Kn_{i}(\frac{1}{n_{i}}\frac{dn_{i}}{dz} - \frac{1}{n(H_{2})}\frac{dn
(H_{2})}{dz}),$ 
where $<>$ denotes the time average of a quantity, and the diffusion coefficient
$K[cm^{2}\;s^{-1}]=<V_{t}L>$ with $V_{t}$ being the velocity for the random 
motion and $L$ the mixing-length scale. 

Now with a transport term, the continuity equation at a cloud position is
$\frac{\partial n_{i}}{\partial t} + \frac{\partial \phi_{i}}{\partial z} = P_{i}
-L_{i}
$
where $P_{i}$ and $L_{i}$ are the chemical production and loss terms, respectively. Defining chemical time-scale $\tau_{i}^{chem}=n_{i}/(P_{i}-L_{i})$ and
diffusion time-scale $\tau_{i}^{diff}=n_{i}/\frac{\partial \phi_{i}}{\partial z}$, the continuity equation reduces to a simpler form
\begin{equation}
\frac{\partial n_{i}}{\partial t}=n_{i}(\frac{1}{\tau_{i}^{chem}}-\frac{1}{\tau_{i}^{diff}}).
\end{equation}

Defining a time constant $\tau=R^{2}/K$ 
which is typically large compared to the chemical time-scales, 
Williams \& Hartquist (1991) argued that diffusion will not be important. 
It is not difficult to see that the real situation is much more complicated 
than this. First,  $\tau=R^{2}/K$ has essentially no physical meaning in the 
formulation. Both the chemical and diffusion time-scales in the above 
continuity equation are position-, time- and
species- dependent. Even when we fix the question to a 
particular moment, both time-scales are still dependent on the detailed physical
and chemical conditions at a cloud position in a complicated manner. 
The diffusion time-scale may be simplified to 
$\tau_{i}^{diff}=H_{i}H/K$, where the scale heights $H_{i}=n_{i}/\frac{\partial n_{i}}{\partial z}$ and $H=f_{i}/\frac{\partial f_{i}}{\partial z}$, but it 
would not come to $\tau=R^{2}/K$ without a large error. Second, 
diffusion is generally coupled to the highly non-linear chemistry and can not
be treated as a pure diffusion in this case;
a small change in some key species due to diffusion between 
neighboring layers can boost the local generation or destruction of certain
species (XAL). 

\section{In the Case of MHD Waves}
It is quite possible that the observed ``turbulence" in molecular clouds is
due to magneto-hydrodynamic waves (MHD waves). 
For a perfectly coherent wave particles 
carried up will be carried down in a cycle, so there will be no net transport.
Would MHD waves be perfectly coherent
in molecular clouds, given the irregularities in magnetic fields, densities and
other physical conditions and even star forming activities ? The magnetic 
Reynolds number is so large that any such waves will be turbulent and random
in nature on relevant spatial scales (Myers \& Khersonsky 1995), which will 
naturally lead to net particle transport (Lindzen 1971; Hunten 1975). 
In fact, even if waves are perfectly coherent, the non-uniform and non-homogeneous chemistry 
will still unavoidably cause a net transport of certain chemical species, as chemicals 
carried up could be destroyed and will thus never be carried down, or vice versa.
Lindzen (1971) suggested that 
the diffusion coefficient in the case of interrupted waves can be expressed 
as $K\sim v\lambda/(\pi x)$, where 
$v$ is the average wave velocity, $\lambda$ is the averaged wavelength and
$x$ is the number of wave cycles (non-integer) before being interrupted. 
Notice that diffusive transport along the average magnetic
field would be enough to have significant effects on the chemistry.

TX's research is supported in part by an NSF grant to the Laboratory for
Millimeter-wave Astronomy at the University of Maryland.

\end{document}